\documentclass{article}
\usepackage[latin1]{inputenc}
\usepackage{amssymb}
\usepackage{amsmath}
\usepackage{amsfonts}
\usepackage{float}
\usepackage[pdftex]{graphicx}
\usepackage[dvips]{hyperref}
\usepackage{capt-of}
\usepackage{times}

\DeclareGraphicsExtensions{.pdf}   
\normalfont

\begin{document}

\begin{abstract}
We discuss some social contagion processes to describe the formation and
spread of radical opinions in the spirit of previous work by Serge Galam.
The dynamics of opinion spread involves local threshold processes as well as
mean field effects. We calculate and observe phase transitions in the
dynamical variables resulting in a rapidly increasing number of passive
supporters. This strongly indicates that military solutions are
inappropriate.
\end{abstract}


\begin{center}
{\huge {Passive Supporters of Terrorism and Phase Transitions} }
\end{center}

Friedrich August$^+$, Philippe Blanchard$^*$, Sascha Delitzscher$^{*,}$%
\footnote{%
foam@physik.uni-bielefeld.de}, Gerald Hiller$^+$, Tyll Krueger$^{*,}$%
\footnote{%
tkrueger@Physik.Uni-Bielefeld.DE}

\begin{center}
\emph{$^+$ Dept of Mathematics, TU Berlin, D-10623 Berlin, Germany}

\emph{$^*$ Dept of Physics, University Bielefeld, D-33615 Bielefeld, Germany}
\end{center}



\section{Introduction}

In this article we discuss some social contagion processes which may play an
important role in the dynamics of radical opinion formation. The prime
applications in mind are conflict situations as they are met at the time of
writing in Afghanistan, Iraq or Palestine where a highly armed alliance of
foreign troops fights against a part of the local population which has been
radicalized in a way such that western social classification dubs them
terrorists. In the following we will use the word "terrorist" solely to
refer to a certain subpopulation (whose interaction features will be
described below) in our model environment and do not intend to enter the
difficult debate of what constitutes the social essence of terrorism. In
terms of our model interaction and shortly speaking one could say that
terrorists are those individuals on which counter terrorist throw bombs on.

Besides the radical terrorist groups there is a much larger grey-area of
supporters of these terrorists. Support has many faces, ranging from just
keeping still about what one knows about terrorists locations or movements
up to supporting terrorists by providing various forms of logistic
infrastructure. Again, we will avoid specifying exactly what is meant with
this notion but use it to describe the potential, predecessor states of an
opinion state from which radical groups may (by whatever means and tools)
recruit new members.

The notion of passive supporters was for the first time used in the
context of mathematical modeling by Serge Galam in \cite{galam},\cite{galam3}%
,\cite{galam4}. He studied the role of passive supporters in a lattice
percolation type model and related terrorism power to the spontaneous
formation of random backbones of people who are sympathetic to terrorism but
without being directly involved in it. His focus was on the correlation
between the number of passive supporters, the physical mobility of active
terrorists and phase transitions associated with it. In a certain sense the
present work is a continuation of Galams work on the subject.We arrive at
similar conclusions as Galam did, namely that for these type of conflicts
military solutions are inappropriate.

In this paper we discuss some aspects of the dynamics of passive support of
terrorist activities in virtual social networks. Our main interest lies in
the study of phase transitions in the number of passive supporters induced
by what is euphemistically called collateral damage as is common as a
consequence of counter terrorist attacks on terrorists moving around in
populated places. Phase transitions in the opinion of large parts of a
population are particularly important since they violate the classical
"linear" action-reaction view common among military leaders and politicians.

We are not concerned with real terrorist networks and their dynamics, for a
recent work on this topic see \cite{Nature} and the references therein. Our
model is based on one paradigm. Counter terrorist strikes lead to collateral
damage. In many cases terrorists live or hide among civilians, and civilian
casualties in turn are likely to cause an increase in the number of passive
supporters\footnote{%
at least in the local family and friendship network of a civilian victim
this is highly plausible} and increase the willingness of civilians to
become members of radical groups.

Cause and effect are not linearly coupled, there is a phase transition
instead. It was widely believed among western observers that in Afghanistan
a large part of the population was supporting the Allied forces at the
beginning of the operation, in spite of many casualties caused by bombings
and other military strikes. Then a change in the public opinion seems to
have occurred, the atmosphere inclined to the disadvantage of the allied
forces and a transition from Allied-friendly to Taliban-friendly took place,
causing a boost in the number of passive supporters.

Passive supporters usually do not reveal their nature towards outsiders and
such phase transitions are hardly observable by Allied forces. The number of
passive supporters can only be measured indirectly via the degree of
cooperation of the civil population. If the fraction of passive supporters
in a population becomes large it is likely that counter terrorists face
increased difficulties to gain help and support from civilians. Further the
recruiting pool for radical organizations can become nearly inexhaustible,
as it has been popularized in News reports about the Gaza Strip. Such a
situation may result in an absurd and tragic solution to secure public
safety and to avoid a downward spiral of violence, like the segregation of
people from each other with a fence.

\begin{figure}[tbp]
\centering 
\includegraphics[width=1.0%
\textwidth]{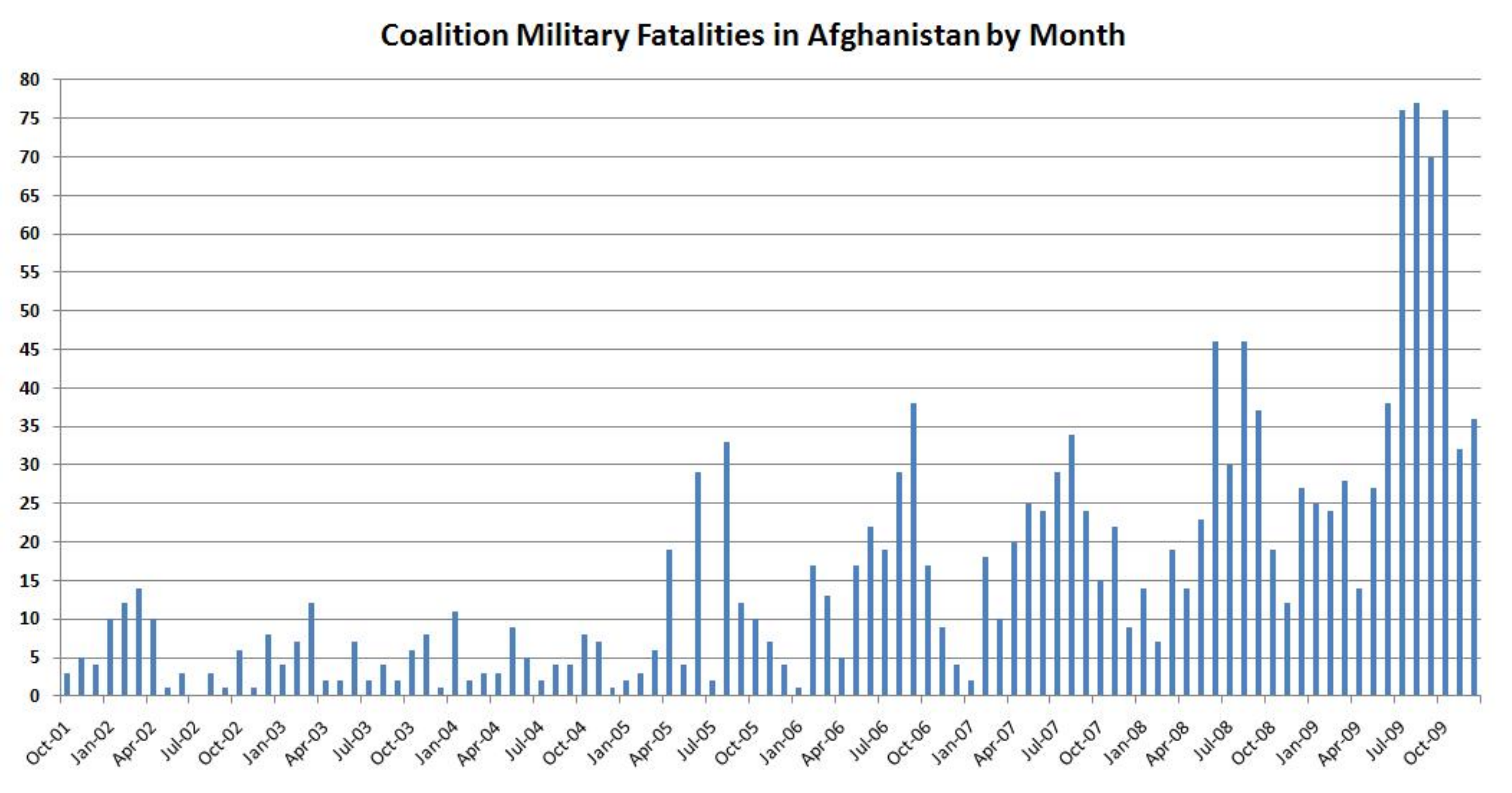}
\caption{Coalition military casualties in Afghanistan by month: Neglecting
seasonal effects, the sudden increase in the number of casualties is very
similar to the density of active terrorists in figures \protect\ref%
{kappa_0_000001_rho1_0_N200000}, \protect\ref{kappa001_rho1} and \protect\ref%
{initinf0_00005_kappa0_0001_t500}, caused by the phase transition in the
density of passive supporters. Source of diagram:
http://www.icasualties.org/oef/}
\label{Coalition_military_casualties_in_afghanistan_by_month}
\end{figure}

\section{Description of the Model}

In the following we will specify the structure of our model which is
inspired by generalized epidemic processes (for more information about this
class of processes and some other applications see \cite{GEP}). Let $G=(V,E)$
be a graph on a finite set of vertices $V,$ representing our model
population, and a set of edges $E$ encoding the social contacts relevant for
transmissions. We start with the description of the state space of the
vertices. To keep things simple we distinguish only four states $S=\left\{
0,1,2,3\right\} $. State $0$ encodes the members of the susceptible
population which are more or less neutral in their opinion about terrorists.
State $1$ individuals correspond to the passive supporters of terrorism and
state $2$ encodes active terrorists. State $3$ vertices correspond to
vertices, which are isolated from the network and do not interact anymore
with other vertices. Although it would be natural, to consider more refined
states we can restrict the discussion of phase transitions to this rather
simple setting by arguing that the type of phase transition described here
is robust to many refinements of the model.

There is usually an immune subpopulation which is resistant to any radical
opinion influence. Including this group will reduce the size of the
susceptible population but does not influence the existence of phase
transitions. Also there usually exist dynamics within the terrorist groups
in complex hierarchical networks which in turn may have a strong influence
on terrorists activities (see \cite{Nature}). These intrinsic network
dynamics within the relatively small subpopulation of terrorists have only
marginal influence on the $0-1$ transition process (the process of becoming
a passive supporter via a social contagion process).

The dynamics will be defined via two transition probabilities $P_{loc}$ and $%
P_{mean}$, where $P_{loc}$ depends on the state distribution in the
neighborhood of a vertex and $P_{mean}$ corresponds to a mean field term
depending on the overall prevalence of the state variables at a given time $%
t $. Although mesoscopic dependencies are in principle possible, we think
they are of secondary importance and do not change the dynamical picture in
general.

With $\chi_i(t)$ denoting the state variable of individual $i$ at time $t$,
for the probability of switching the state from $a$ to $b$ we formally have:

\begin{equation*}
P\{\chi _{i}(t+1)=b\mid \chi _{i}(t)=a\}=P_{loc}\left( \chi
_{B_{1}(i)}(t),b\right) \oplus P_{mean}\left( s(t),a,b\right) ;a,b\in
\left\{ 0,1,2,3\right\}
\end{equation*}%
where the sum sign stands for independent superposition\footnote{%
Independent superposition of two probabilities $A$ and $B$ is just $A+B-AB$}
of the local and global infection events. The subscript $B_{1}(i)$ denotes
the neighborhood of $i$ and $s(t)$ the total number (or density) of vertices
not in a susceptible or immune state at time $t$.

The local dynamics we use here is inspired by generalized epidemic processes 
\cite{GEP}. In classical epidemics, the probability of infection is an
independent superposition of the single infection events caused by each
infected neighbor (for small infection probabilities this means that the
probability to get infected is proportional to the number of infected
neighbors). In generalized epidemic processes to every vertex $i$ is
assigned a threshold $\Delta _{i}$ of infected neighbors, at which the
probability of infection suddenly jumps to a high probability. More
precisely, let $N_{i}(t)$ denote the number of neighbors of a susceptible
individual $i$ at time $t$ which support or accomplish terrorism. Hence $%
N_{i}(t)$ is the sum of state $1$ and state $2$ neighbors of $i$ at time $t.$
Then, the probability that individual $i$ will become a passive supporter at
time $t+1$ is given by:

\begin{equation*}
P_{loc}\left\{ \chi _{i}(t+1)=1\mid \chi _{i}(t)=0\right\} =\left\{ 
\begin{array}{ll}
\epsilon N_{i}(t), & N_{i}(t)<\Delta _{i} \\ 
\alpha , & N_{i}(t)\geq \Delta _{i}%
\end{array}%
\right.
\end{equation*}

This type of local dynamics applies with appropriate modifications to many
social processes like the spread of rumors, prejudices, knowledge or
beliefs. So, whenever one vertex has contact with too many (here $\Delta $
or more) passive supporters or terrorists, they will sooner or later become
passive supporters of the terrorists, either because they get convinced by
the ideology or because they get exploited, unintentionally or even unnoted
by themselves. Passive supporters as well as terrorists themselves may
become uncooperative or useless, so they discontinue support. We assume
therefore that for every vertex there is a chance of switching back from $1$
or $2$ to $0$ with a given probability $\gamma $.

Also there is a fixed rate $\kappa$ at which terrorists recruit from passive
supporters. Due to the action of external forces like the state or allied
troops, or by death or by migration, terrorists can be neutralized and
removed from the network with probability $\rho$. Emotionally involved
relatives and friends of victims, who where in state $0$ until then, will
potentially decide to join the terrorists side to take revenge. Civilians
watching the News may be upset and enraged when they observe counter
terrorists taking out their fellow citizens and destroying their buildings
to capture or to eliminate terrorists.

We assume a fixed rate $\beta$ at which a neutralized terrorist induces new
passive supporters via the previously described social processes. Thus, for
a susceptible vertex in the network, the probability of becoming a passive
supporter depends (besides the above describe local infection way) on the
number of neutralized terrorist per time step and therefore on the number of
active terrorists in the network. In this way, $\beta$ constitutes the mean
field dynamics of the infection process. If $C(t)$ denotes the number of
captured terrorists at time $t$ we have formally for the probability that a
state $0$ vertex changes its state into $1$ in the next time step:

\begin{eqnarray}
P_{mean}\left\{ \chi _{i}(t+1)=1\mid \chi _{i}(t)=0\right\} &=&1-\left(
1-\beta \right) ^{C\left( t\right) } \\
&\sim &\beta C\left( t\right) ;\text{for small }\beta.
\end{eqnarray}

Note that if one ignores mass media effects the total number of new mean
field induced passive supporters should stay bounded irrespective of the
size of the population (reflecting the fact that it is at most an effect on
mesoscopic scale). We therefore scale the $\beta$ value with the population
size appropriately.

Table \ref{parametertable} summarizes all parameters and their
interpretations.

\begin{table}[h]
\caption{Summary of all parameters}
\label{parametertable}%
\begin{tabular}{c|c|l}
parameter & transition & affiliation \\ \hline
$\epsilon$ & $0\rightarrow1$ & local process below the threshold \\ 
$\alpha$ & $0\rightarrow1$ & local process above the threshold \\ 
$\beta$ & $0\rightarrow1$ & mean field \\ 
$\gamma$ & $1,2\rightarrow0$ & changing mind \\ 
$\kappa$ & $1\rightarrow2$ & procreation rate of new terrorists \\ 
$\rho$ & $/$ & neutralization rate of terrorists%
\end{tabular}%
\end{table}

\section{Phase Transitions in the Local Infection Process}

The usage of local threshold dynamics yields additional phase transitions
that are different from those in classical epidemics, which are phase
transitions in the parameters. Here, in addition to these phase transitions
in the parameters there are such in the dynamical variables like the number
of passive supporters. In the following we state some analytic results for
these phase transitions for the case of the classical Erd�s\&Renyi random
graph $G(n,p),$ the graph with $n$ vertices and independent edge probability 
$p$ between each pair of vertices.

To investigate the average infection density $b_{t}$ on a $G(n,p)$ with $%
p=c/n$ we assume for simplicity $\epsilon =0,\alpha =1$ and that the
thresholds $\Delta _{i}$ are distributed uniformly, i.e. $\Delta _{i}=\Delta 
$ for all vertices $i$. It is easy to see that the case of very small $%
\epsilon$ and rather large $\alpha $ gives similar results although the
formulas become more complicated. For a detailed mathematical treatment see 
\cite{GEP}. There is a recursion for $b_{t}$, namely

\begin{equation*}
b_{t+1}=1-(1-b_{0})e^{-cb_{t}}\sum_{k=0}^{\Delta -1}\frac{(cb_{t})^{k}}{k!}.
\end{equation*}

Thus, for $\Delta=2$ we get for the fixed point equation

\begin{equation}
b^{\ast }=1-\left( 1-b_{0}\right) e^{-cb^{\ast }}\left( 1+cb^{\ast }\right).
\label{fixedpt}
\end{equation}%
Note that there can be several solutions one has to take the dynamically
stable one (closest to $b_{0}$ ). A closer examination shows that, depending
on the value of $b_{0}$ and $c,$ there are either 3 fixed points or just one
in the domain $[0,1]$. Since the iteration mapping has no critical points in
this interval the smallest of these fixed point is also the attractor for
the orbit starting with $b_{0}.$ Three examples are shown below. The phase
transition happens when the first two fixed points (in case there are three)
join together and form an indifferent (slope one) fixed point, see fig. \ref%
{fixedpt}. That means that for $b_{0}<b_{0}^{c}$, i.e. if the initial
prevalence is smaller than the critical value, the final infection density $%
b^{\ast }$ will be not much larger than $b_{0}$. If $b_{0}>b_{0}^{c}$ then $%
b^{\ast }$ will be close to the number of vertices in the $k$-core\footnote{%
A subgraph $G^{\prime }$ of a graph $G$ is called $k$-core of $G$ if every
vertex in $G^{\prime }$ has at least $k$ neighbors.} of the graph. The phase
transitions are closely connected with the k-cores of a network. If $k\geq
\Delta $ then every vertex in the k-core succeeds the threshold and gets
infected sooner or later.

The critical density $b_0^c$, depending on the edge density $c$ of the
graph, is given by

\begin{equation*}
b_0^c=1-\frac{2\exp(-\frac{1}{2}(1-c+\sqrt{c^2-2c-3}))}{c(-1+c-\sqrt{c^2-2c-3%
})}
\end{equation*}

\begin{figure}[]
\centering 
\includegraphics[scale=0.35]{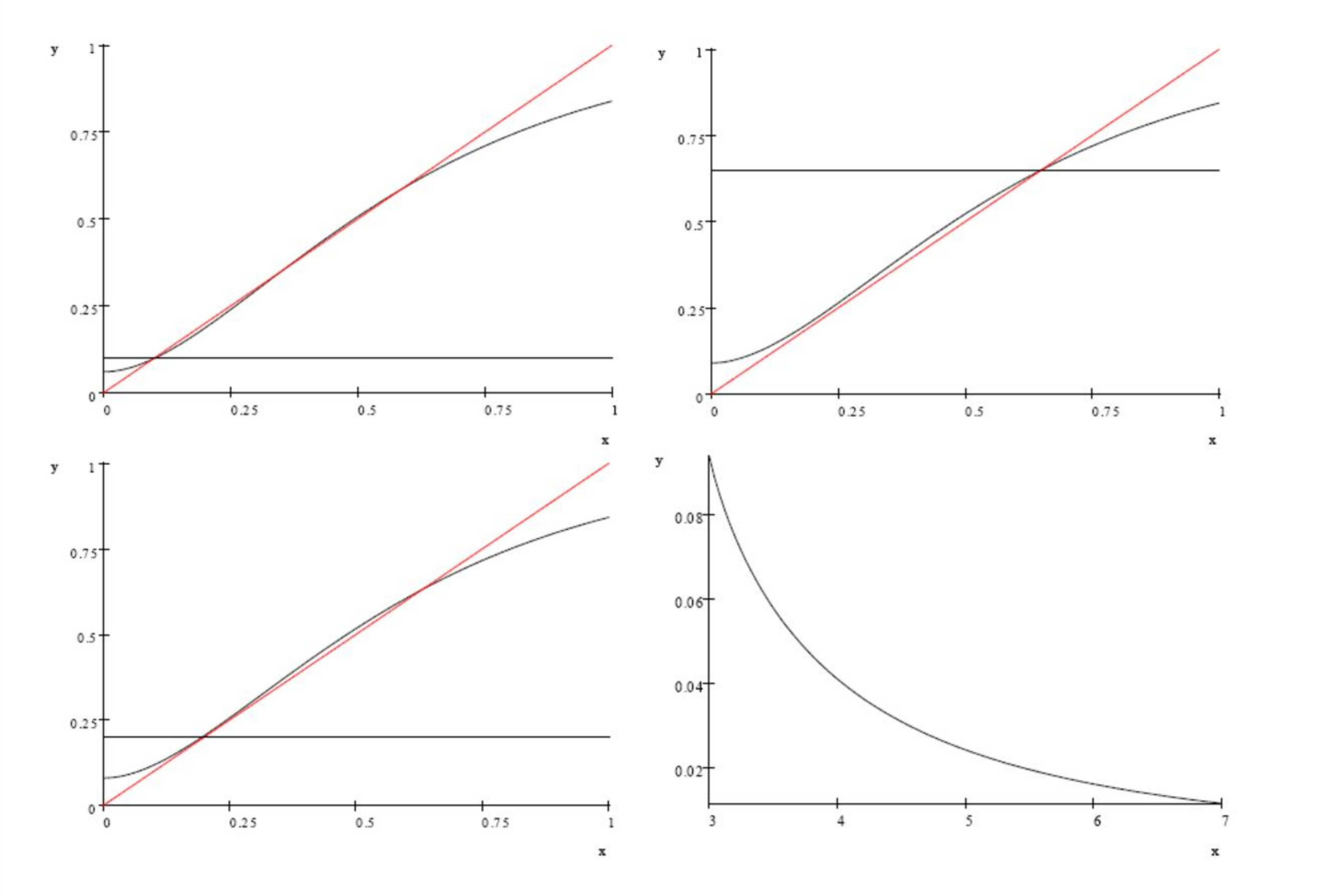}
\caption{Right hand side of eq \protect\ref{fixedpt} (top left, top right,
lower left). The black lines point to the stable fixed point. The graph in
the lower right shows $b_0^c$ as a function of $b_0$.}
\label{phasetransition}
\end{figure}

Figure \ref{phasetransition} shows the function of the right hand side of
eq. \ref{fixedpt}.

The mean field dynamics described in section $2$ induces now an effective
slow increase of the initial density for the local spreading process until
the critical density for the local dynamics is reached. Above the phase
transition there remains then a nearly inexhaustible pool for recruiting new
terrorists and any attempt to resolve the situation by military means has to
fail. We close this section with a short calculation for the pure mean field
dynamics. Let $\kappa$ be the probability that a passive supporter becomes a
terrorist (for the simulations this is taken to be $0.1$ percent). Assume
that $n$ is large and $\beta n=B$, Then the branching process approximation
to the pure mean field process (which is valid for small numbers of
initially infected) becomes overcritical if 
\begin{equation}
\frac{\kappa\rho B}{\left(\gamma
+\kappa(1-\gamma)\right)\left(\gamma+\rho(1-\gamma)\right)} > 1.
\end{equation}%
Note that in case the critical threshold for the local infection process is
very small even in the subcritical case the mean field dynamics can due to
stochastic fluctuations trigger the dynamics above to the phase transition.

\section{Simulation Results}

The simulations are run on an Erd�s\&Renyi random graph $G(n,p)$ in discrete
time and synchronous updates. Throughout all simulations the parameters of
the local infection process are chosen to be $\alpha =1$, $\epsilon =0$.
Note that the difference to a process with transmission probability $\alpha
=0.5$ would be just a small time delay. The remaining parameters are chosen
to be plausible if one interprets the time step to be one week with the
situation in Afghanistan in mind.

On average, one percent of the passive supporters change back to being
susceptible, $\gamma =0.01$. The probability of becoming an active terrorist
is zero for all susceptible vertices and $0.1$ percent for all passive
supporters. Thus, on average $1000$ passive supporters generate one single
terrorist.

On each picture the abscissa is time, where we - as was mentioned above -
interpret one time step $\Delta t=1$ to be one week. There are other
associations possible of course, but in the context of social networks in
regions, which are struck by terror, associating one time step with one
week, giving people the opportunity to act and communicate, seems to be a
natural choice.

The y-axis is the density of vertices in state $1$ or $2$, thus having
values running from $0$ to $1$. There are three graphs shown, the dark blue
one being the number of active terrorists, the dark green one being the
number of passive supporters and the turquoise one being the sum of these
two.

\begin{figure}[]
\begin{minipage}{6cm}
      \centering 
      \includegraphics[scale=0.27]{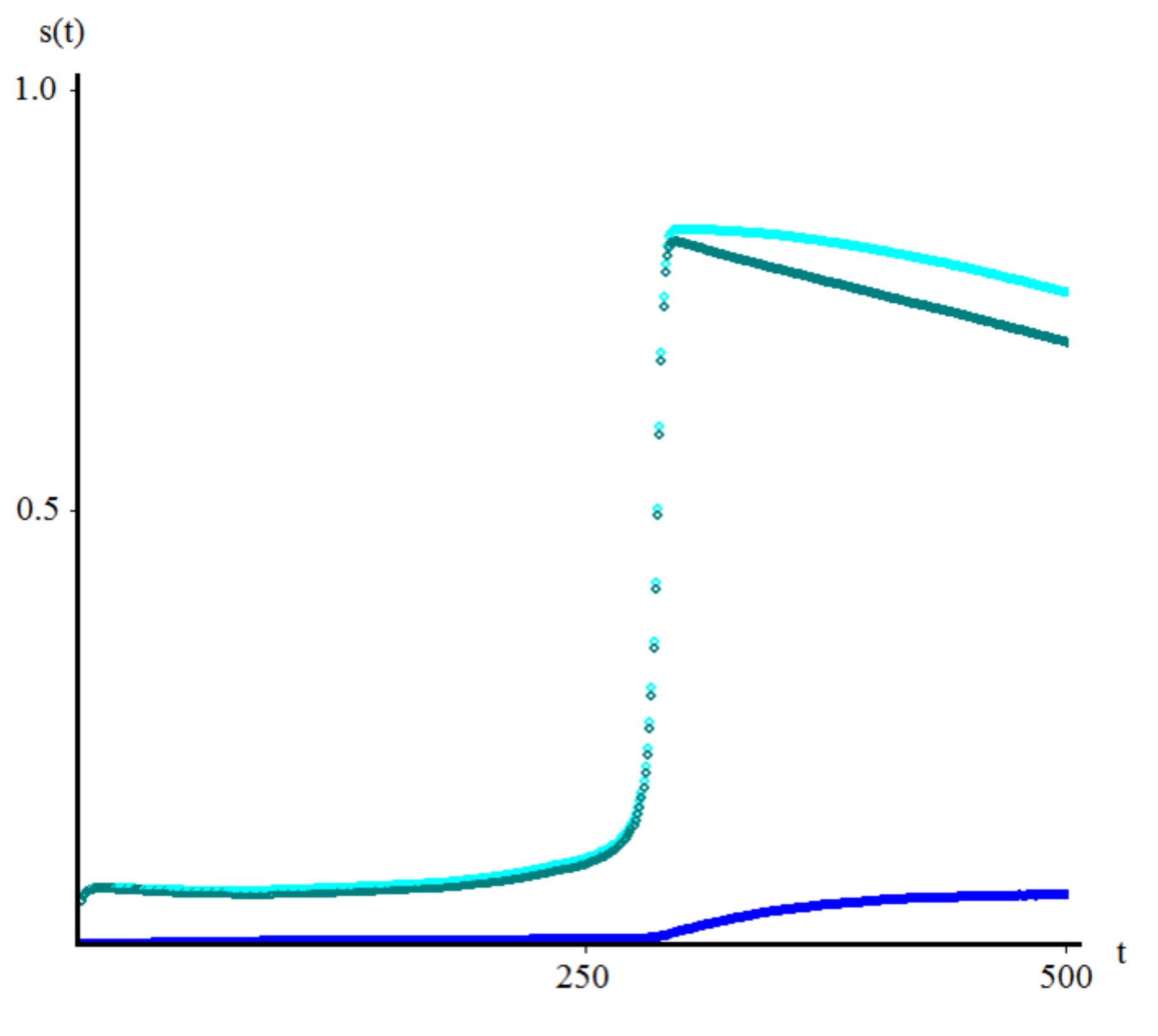}
   \end{minipage}
\begin{minipage}{6cm}
      \centering 
      \includegraphics[scale=0.27]{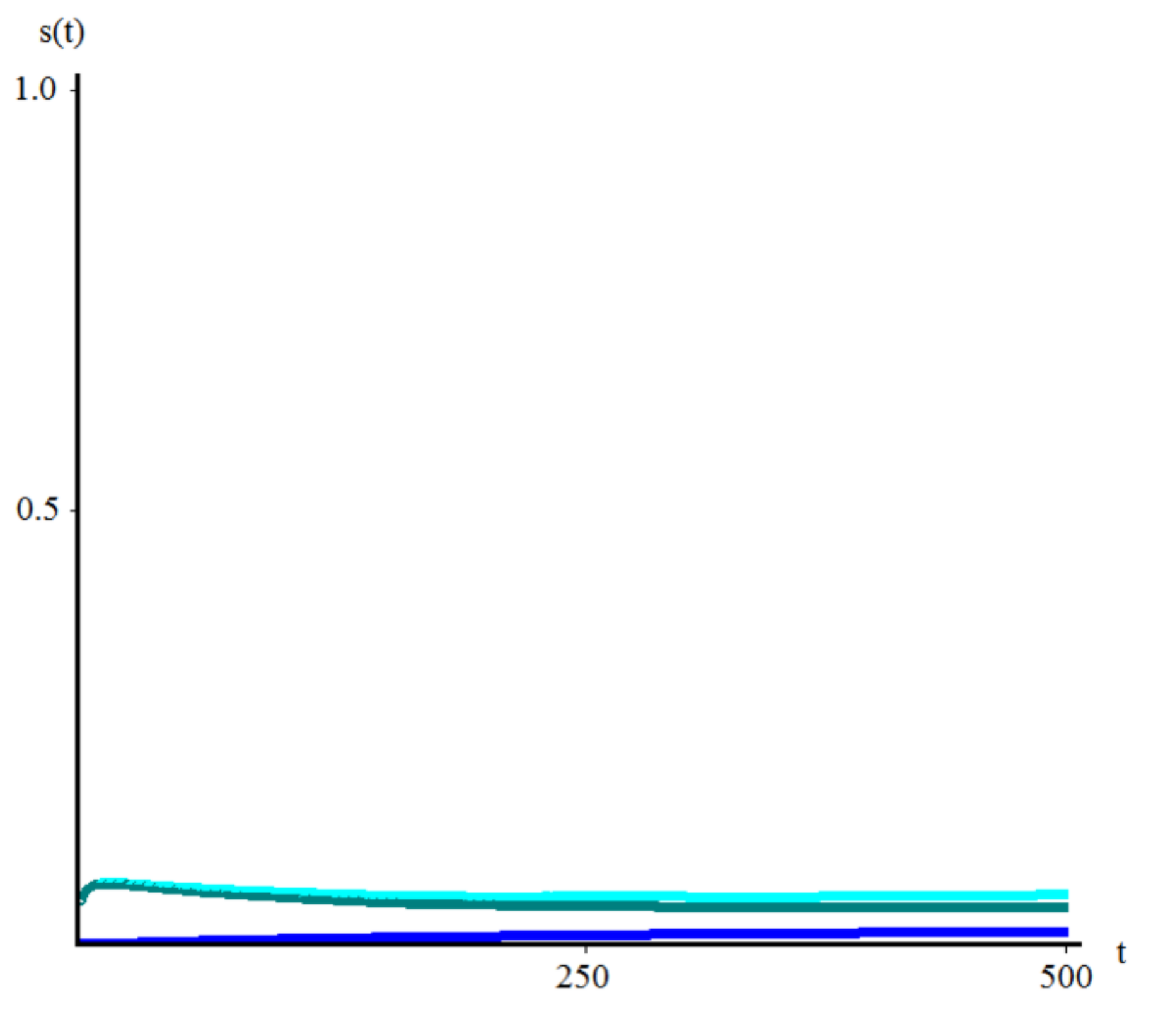}
   \end{minipage}
\caption{Densities of active terrorists and passive supporters on an Erd�%
s-Renyi random graph $G(n,p), p=c/n$ with $n=200^{\prime}000$ and $c=4$,
threshold $\Delta=2$, initial density of passive supporters $b_0=0.04$, mean
field $\protect\beta=0.00001$, in a time span of $t=500$ steps. Choosing $%
\protect\rho=0.01$ (left) yields a phase transition in the number of passive
supporters and a sudden growth of the number of active terrorists, while
choosing $\protect\rho=0.001$ (right) yields no phase transition or a phase
transition outside of the time scale under consideration.}
\label{kappa_0_000001_rho1_0_N200000}
\end{figure}

Figure \ref{kappa_0_000001_rho1_0_N200000} shows the case of an Erd�s-Renyi
random graph $G(n,p), p=c/n$ with $n=200^{\prime}000$ and and average degree
of $c=4$. The threshold for the local dynamics is set to $\Delta=2$, i.e. as
soon as one vertex has two passive supporters in its neighborhood it turns
to a passive supporter as well. The initial density of passive supporters is
set to $b_0=0.04$, i.e. there are $8000$ passive supporters from the start.
The mean field parameter $\beta$, which is the rate at which one captured
active terrorist generates passive supporters, is set to $\beta=0.00001$,
i.e. one captured active terrorist generates on average 2 passive supporters.

The process is run for a span of $t=500$ steps, corresponding to about 10
years of time. Choosing $\rho=0.01$ (left) yields a phase transition in the
number of passive supporters and a sudden growth of the number of active
terrorists, while choosing $\rho=0.001$ (right) yields no phase transition.
While in both cases the number of passive supporters increases directly from
the start to a level, which is slightly decreasing for one third of the
time, the higher value of the capturing rate triggers the mean field process
to produce enough passive supporters to lift the local process to an
overcritical level, where the calamitous phase transition occurs, resulting
in an irreversible increase of active terrorists.

The impact of changing the capturing parameter $\rho$ is shown in Figure \ref%
{kappa001_rho1}, on an Erd�s\&Renyi random graph $G(n,p), p=c/n$ with $%
n=50^{\prime}000$ vertices and average degree $c=4$, threshold $\Delta=2$,
initial density of passive supporters $b_0=0.01$, i.e. $500$ passive
supporters from the start. The mean field parameter is $\beta=0.001$, i.e.
one captured active terrorist generates $50$ passive supporters here.
Choosing $\rho=0.004$ (left) or $\rho=0.0025$ (right) only alters the moment
when the phase transition happens.

\begin{figure}[]
\begin{minipage}{6cm}
      \centering 
      \includegraphics[scale=0.27]{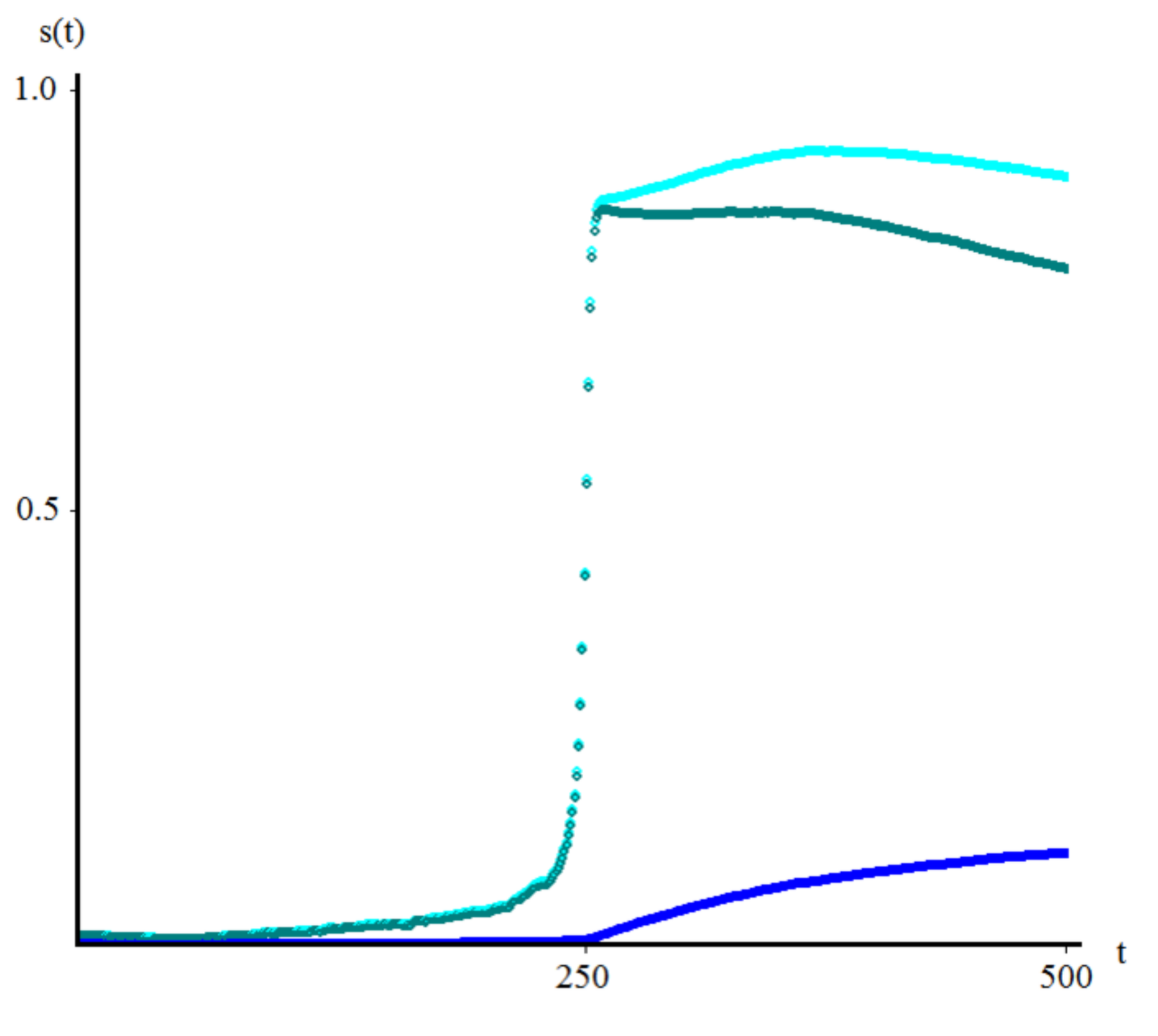}
   \end{minipage}
\begin{minipage}{6cm}
      \centering 
      \includegraphics[scale=0.27]{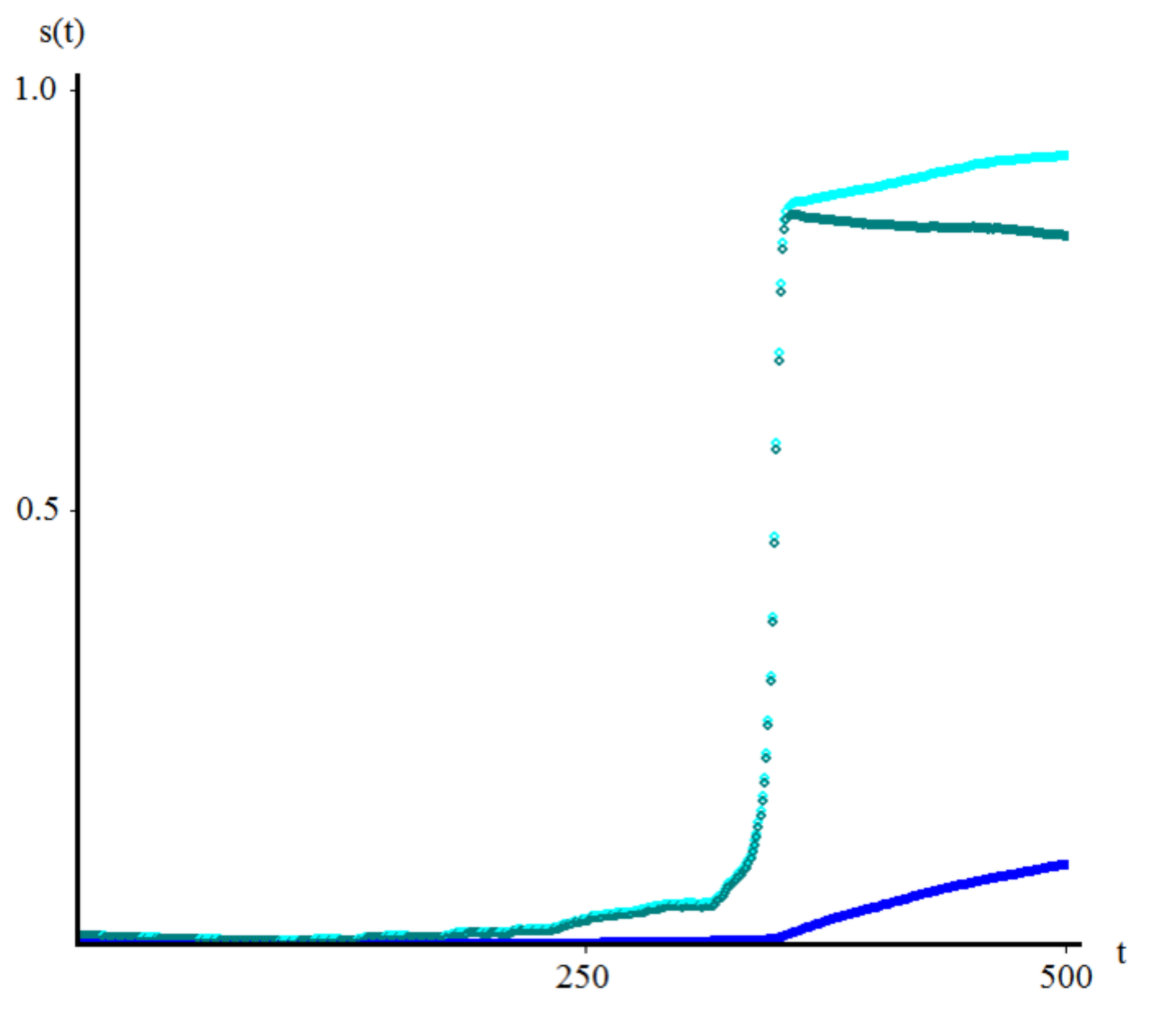}
   \end{minipage}
\caption{Densities of active terrorists and passive supporters on an Erd�
s-Renyi random graph $G(n,p), p=c/n$ with $n=50^{\prime}000$ and $c=4$,
threshold $\Delta=2$, initial density of passive supporters $b_0=0.01$, mean
field $\protect\beta=0.001$, in a time span of $t=1500$ steps. Choosing $%
\protect\rho=0.004$ (left) or $\protect\rho=0.0025$ (right) only alters the
moment when the phase transition happens.}
\label{kappa001_rho1}
\end{figure}

\begin{figure}[]
\begin{minipage}{6cm}
      \centering 
      \includegraphics[scale=0.27]{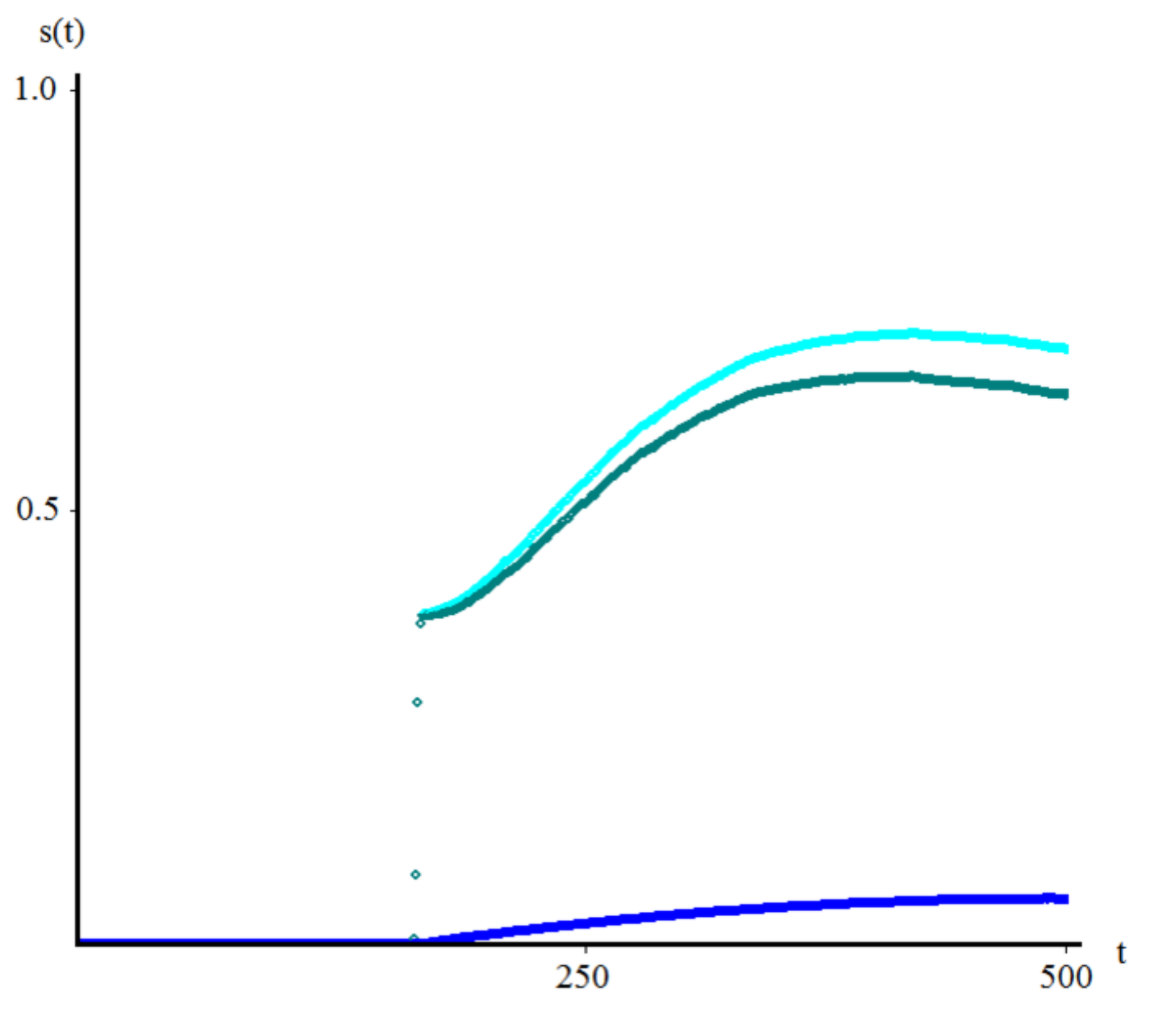}
   \end{minipage}
\begin{minipage}{6cm}
      \centering 
      \includegraphics[scale=0.27]{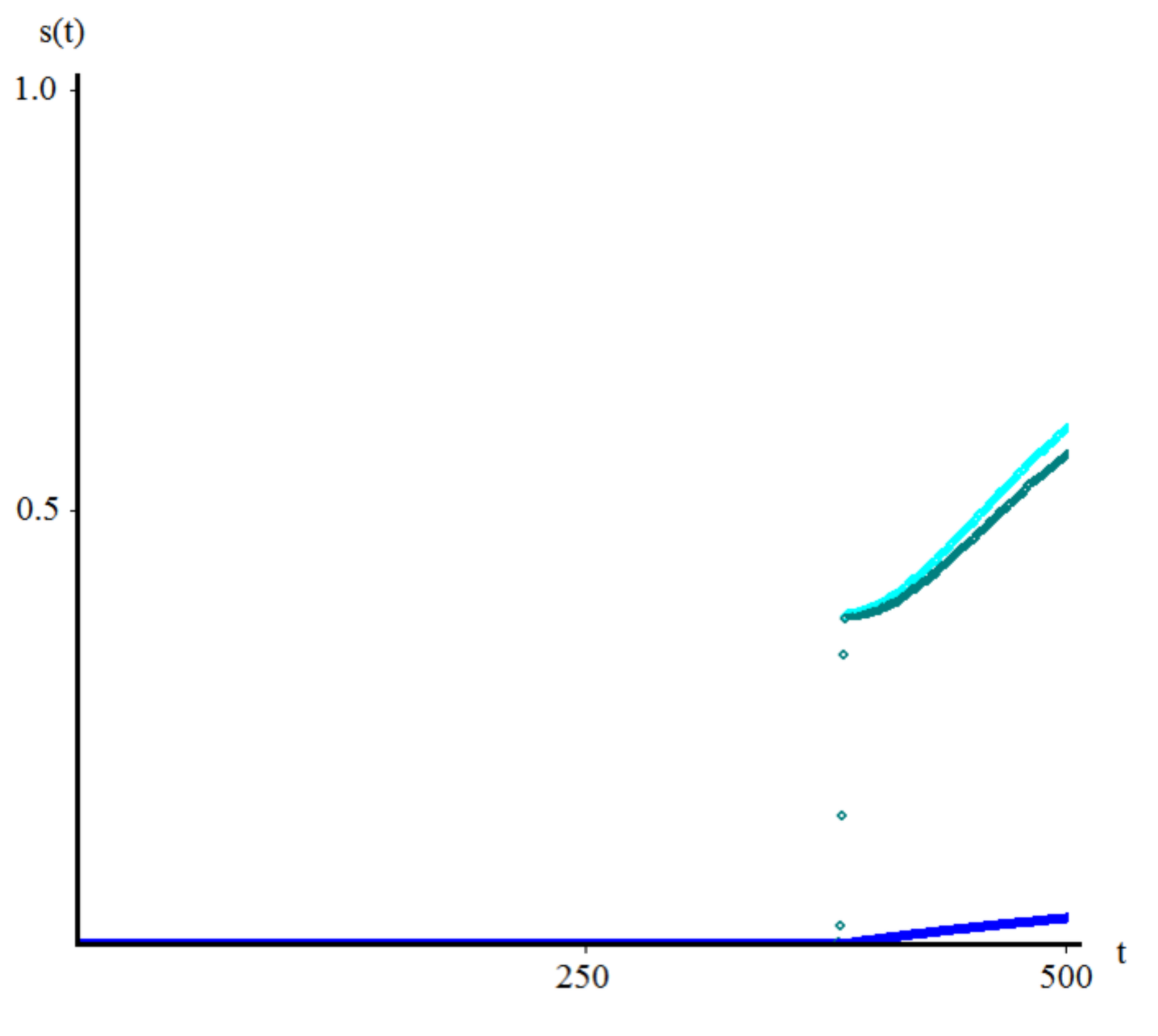}
   \end{minipage}
\caption{Densities of active terrorists and passive supporters on a sample
of the StudiVz network. The threshold is $\Delta=2$, initial density of
passive supporters $b_0=0.00005$, mean field $\protect\beta=0.0001$, in a
time span of $t=500$ steps. The phase transition occurs as soon as one or
two active terrorists get captured.}
\label{initinf0_00005_kappa0_0001_t500}
\end{figure}

Fig \ref{initinf0_00005_kappa0_0001_t500} shows the simulation run on a
sample of the anonymized StudiVz network, including members associated with
scholars of Bielefeld and their friends. The network has a size of
approximately $400^{\prime}000$ vertices with an average degree of about $%
7.2 $, for more details see \cite{conradlee}. With an initial density of
passive supporters of $b_0=0.00005$, corresponding to an average of $20$
passive supporters from the start, and with capturing rate $\beta=0.0001$,
corresponding to an average of $40$ citizens getting converted to passive
supporters, we observe the phase transition occurring as soon as one or two
active terrorists get caught.

\section{Conclusions and Perspectives}

The main observation is the existence of a phase transition in the number of
passive supporters of terroristic activities. Whenever counter terrorist
activities lead to collateral damages, the likelihood of outraging civilians
rises. A high number of passive supporters provides a steady pool to recruit
active terrorists, so the number of active terrorists and their attacks
increases, as fig. \ref%
{Coalition_military_casualties_in_afghanistan_by_month} suggests and as
resembled by the graph for the number of active terrorists in figures \ref%
{kappa_0_000001_rho1_0_N200000}, \ref{kappa001_rho1} and \ref%
{initinf0_00005_kappa0_0001_t500}.

Our results not only suggest lowering of the rate $\rho $ of removal of
active terrorists to avoid the phase transition. The interplay of the mean
field term $\beta$, which is the rate at which removed active terrorists
generate passive supporters, and $\rho $ has to be taken into account.
Avoidable failures resulting in casualties, high collateral damage, pictures
and videos of humiliated inmates in Allied prisons, are factors which
increase the probability that the civil population will join the terrorist
side instead of fighting against it.

If the Allied forces want to avoid the phase transition in the number of
passive supporters to not gain a stable number of active terrorist,
capturing or removing active terrorists from the network would make sense
therefore only if this happened practically without casualties, fatalities,
applying torture or committing terroristic acts against the local population.

If this is not possible - and evidence is pointing towards this - our
results strongly indicate that there is no military solution to fight
terrorism, so only political solutions are available, in strict
coincidence with former results obtained by Galam, \cite{galam3},\cite%
{galam4}.

A refinement of the model may use weighted graphs, where the vertices have
properties like credibility or importance, to include intra-organization
dynamics, e.g. to model the emergence of terror cells. In the same fashion
the influence of local warlords and tribal conflicts may be described.
Finally, it will be important to extend the model to allow studying also the
countereffects of the terrorists themselves in case their actions produce
more and more civilian victims.

\end{document}